\documentclass[final,5p,times,twocolumn]{elsarticle}
\usepackage{epsfig}
\usepackage{ae}
\usepackage{hyperref}
\usepackage{lineno}
\usepackage{graphicx}
\usepackage{amssymb}
\usepackage{amsmath}
\newcommand{\nima}{Nucl. Instr. and Meth. A}
\begin{document}

\begin{frontmatter}



%\title{Type the title of your paper, capitalize first word}
% \title{Development of a large size GEM Chamber and testing at high rate of proton beam}
\title{Performance of a large size triple GEM detector at high particle rate for the CBM Experiment at FAIR}
%%%%%\title{Most probable energy loss as a function of the drift distance}
%\title {An estimation of the effective number of electrons \\
%           contributing to the coordinate measurement with a TPC \\
%                             II}
%\author[label1,label2]{}

%
%%%%%\author{Makoto Kobayashi}
%
%%%%%\address{High Energy Accelerator Research Organization (KEK),
%%%%%                                         Tsukuba, 305-0801, Japan}
%
\iffalse   \author[label1]{Rama Prasad Adak},
   % \corref{mycorrespondingauthor}},
%    \cortext[mycorrespondingauthor]{Corresponding author}
%    \ead{rpadak@jcbose.ac.in}
\author[label2]{Ajit Kumar}
\author[label2]{Anand Kumar Dubey}
\author[label2,label1]{Subhasis Chattopadhyay}
\author[label1]{Supriya Das}
\author[label1]{Sibaji Raha}
\author[label1]{Subhasis Samanta}
\author[label2]{Jogender Saini}

%
%***************
%   Addresses
%***************
%
\address[label1]{Center for Astroparticle Physics $\&$ Space Science, Block EN, Sector V, Saltlake, Kolkata, 700091, India and Department of Physics, Bose Institute, 93/1, A.P.C. Road, Kolkata 700009, India}
\address[label2]{Variable Energy Cyclotron Centre, Kolkata, India}

\fi
% \input{authorlist.tex}
\author[1]{Rama Prasad Adak}
\ead{rpadak@jcbose.ac.in}
\author[2]{Ajit Kumar}
\author[2]{A. K. Dubey}
\author[1]{Subhasis Samanta}
\author[2]{J. Saini}
\author[1]{S. Das}
\author[1]{S. Raha}
\author[2,1]{Subhasis Chattopadhyay}

%
%***************
%   Addresses
%***************
%
\address[1]{Centre for Astroparticle Physics \& Space Science, Bose Institute, Block EN, Sector V, Salt Lake, Kolkata 700091, India and Department of Physics, Bose Institute, 93/1, A.P.C. Road, Kolkata 700009, India}

\address[2]{Variable Energy Cyclotron Centre, Sector-1, Block-AF, Salt Lake, Kolkata, India}

%%%\linenumbers

%
%===============================================================
%
%   Abstract
%
%===============================================================
%s
\begin{abstract}
In CBM Experiment at FAIR, dimuons will be detected by a Muon Chamber (MUCH) consisting of segmented absorbers of varying widths 
and tracking chambers sandwiched between the absorber-pairs. In this fixed target heavy-ion collision experiment, operating at highest interaction 
rate of $10~MHz$ for $Au+Au$ collision, after the first MUCH detector station in its inner radial ring will face a particle rate of $1~MHz/cm^2$. To operate at 
such a high particle density, GEM technology based detectors have been selected for the first two stations of MUCH. We have 
reported earlier the performance of several small-size GEM detector prototypes built at VECC for use in MUCH. In this work, we report on a large GEM 
chamber prototype  tested with proton beam of momentum $2.36~GeV/c$ at COSY-J\"{u}elich Germany. The detector was read out 
using nXYTER ASIC operated in self-triggering mode. An efficiency higher than $96\%$ at $\Delta V_{GEM}~=~375.2~V$ 
was achieved. The variation of efficiency with the rate of incoming protons has been found to vary 
within $2\%$ when tested up to a maximum rate of $2.8~MHz/cm^2$. The gain was 
found to be stable at high particle rate with a maximum variation of $\sim~9\%$.

\end{abstract}
\begin{keyword}
%% keywords here, in the form: keyword \sep ke yword

%Lead/scintillator sampling calorimeter \sep
%Wavelength-shifting fiber

Micro-pattern Gas Detector\sep
CBM\sep
Gas Electron Multiplier\sep
Triple GEM detector\sep
MUCH



\end{keyword}

\end{frontmatter}

\section{Introduction}
\begin{figure}
      \begin{center}
	\includegraphics[width=80mm, height = 50 mm]{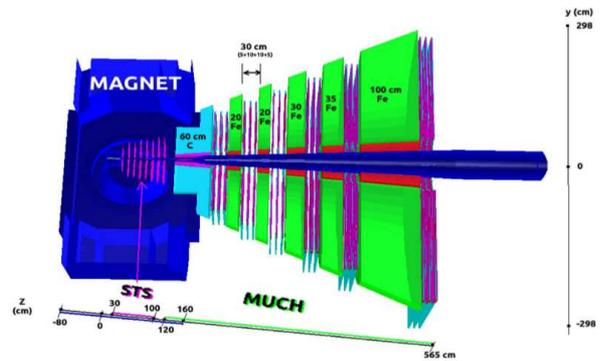}
      \end{center}
	\caption{A schematic view of the MUCH of CBM Experiment with first absorber as carbon and rest are iron.}
	\label{fig:much}
  \end{figure}

The Compressed Baryonic Matter experiment at FAIR \cite{cbm} will explore the region of the phase diagram of strongly interacting matter 
at high baryon density and moderate temperature. This fixed target experiment will use proton to Au ions beams with maximum energy per nucleon, $E_{lab}/A$ of 
$90~AGeV$ for protons and $35~AGeV$ for Au-ions colliding with various target nuclei. The experiment aims to study the chiral symmetry restoration, search 
for the phase transition, locate the critical end point, study the equation of state at high baryon density among other topics. The observables 
of this experiment include low mass vector mesons (LMVMs) like $\rho, \omega$, charmonia along with the collective flow 
of particles, their correlations and fluctuations. The main challenges include the measurement of low multiplicity, rare probes with high accuracy. 
In order to attain reasonable statistics for rare probes at a reasonable running period, the interaction rate of colliding ions in this 
experiment will reach $\sim10~MHz$. In CBM, main tracking device is a set of silicon tracking stations (STS) placed inside a dipole magnet. 
The system measures the momentum of charged tracks with a resolution ($\delta p/p$ of 1$\%$). The LMVM like $\rho, \omega, \phi$ and 
charmonia will be reconstructed from their decay into dileptons. The CBM Muon Chambers (MUCH) consists of alternating layers of hadron 
absorbers and detector stations to track muons. These segmented absorbers allow to identify muons over a wide range of momentum depending 
on the number of segments it passes. A schematic layout of MUCH is shown in Fig.~\ref{fig:much}.
MUCH will cover an acceptance from $\pm~5.6^{\circ}$ to $\pm~25^{\circ}$. The minimum value of the acceptance is the acceptance cover by the beam-pipe whereas 
the maximum value is the opening of the dipole magnet. MUCH 
will be operated in different setup options by varying the positions of the absorber-detector 
combinations. The combinations include $3,~4,~5$ or 6 such pairs for use in SIS$100$ and SIS$300$ energy regions of FAIR 
and two measurement options i.e., LMVM and charmonia. 
% The innermost part of the first detector (i.e. for small polar emission angles) will face a particle density up to 
% $0.4~MHz/cm^{2}$ as obtained from FLUKA simulation\cite{fluka} for $Au+Au$ collision at $E_{lab}=35~AGeV$. 
The first detector station of MUCH will have to face a hit density of $0.1~/cm^2/event$ corresponding to $1~MHz/cm^{2}$ for interaction rate of $10~MHz$ of the colliding ions as obtained from 
 GEANT3 simulations\cite{geant3} using particles from UrQMD event generator\cite{urqmd} for central Au$+$Au collisions at $E_{lab}=25~AGeV$. The choice of the detector technology 
is guided by the rate capabilities of the detectors coupled to the cost to cover a large area. Considering the detector 
technologies presently available or under intense research and developing phase, Gas Electron Multiplier (GEM) \cite{sauli} technology based chamber is found to be a suitable 
candidate. GEM based detectors has been used in CMS, COMPASS, PHENIX for their excellent rate handling capacity \cite{highrate,bressan,bachmann}. ALICE experiment 
will use triple-GEM detectors for their TPC upgrade to handle high rate of particles. 
CBM will use GEM chambers as their tracking detectors in the first two stations of the muon detection system. 
Towards this goal, we at VECC-India have built several triple GEM chambers of dimensions $10 ~cm \times 10~ cm$ and $31~ cm \times 31~ cm$. 
These chambers have been tested with X-rays \cite{cbm10}, proton  \cite{cbmreport13} and pion beams \cite{dubey} to 
achieve $>$ 95\% efficiency. Even after the successful development of these chambers, testing of two main criteria of their use 
in CBM-MUCH remain unfulfilled i.e., large size and high rate capability.

In this work, we report the development of a sector-shaped triple-GEM chamber of 80 cm length and 40 cm longer width. 
We also report the performance of the chamber using proton beams of momentum $2.36~GeV/c$ at the highest rate of $2.8~MHz/cm^2$. 
The present chamber is considered as a prototype for the first station downstream of the magnet that faces the highest 
particle density. The 1$^{st}$ station of CBM will have 3 layers with 16 sector-shaped chambers in each layer.

The paper is organized as follows, in the next section, we discuss in somewhat details the layout of the GEM chambers 
for the CBM-MUCH. Section-3 contains the fabrication procedure of the GEM chamber including details of various components 
followed by the test setup, results and discussions in section 4 and 5 respectively. %We have discussed the results in section 6.

\section{Layout of chambers in CBM muon system}

\begin{figure}
      \begin{center}
 	\includegraphics[width=80mm, height = 60 mm]{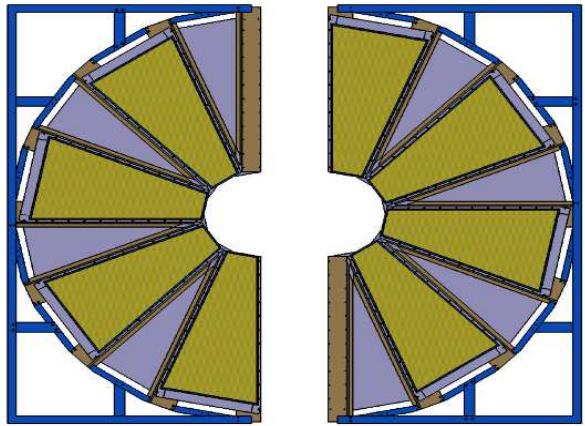}
      \end{center}
	\caption{layout of sectors on a layer.}
	\label{cbm-sector}
  \end{figure}

Fig.~\ref{cbm-sector} shows the schematic diagram of a layer consisting of the sector-shaped chambers. Three such layers are to be mounted in 
a 30 cm gap between two successive absorbers. The number of sectors in each layer for the 1$^{st}$ and 2$^{nd}$ stations are 16 and 24 
respectively. There will be a provision for a layer to be separated into two halves for servicing. For ease of production, all chambers in a 
particular station are identical.
The sector-shaped chamber will be mounted back to back on two planes separated by an Aluminium plate. The active area of each sector will be 
somewhat larger than the area corresponding to 360$^\circ$ divided by the number of sectors. A single GEM chamber in the first station will 
cover $23^{\circ}$ on azimuth including the overlap. This facilitates the overlap at the edges between two sectors. There are overlaps 
of $0.5^\circ$ that corresponds to the mechanical support at the sides of the chambers. 

\section{Fabrication of the GEM chamber}
Present chamber being discussed is a real size prototype chamber for the first station of MUCH. The design and the fabrication of 
readout PCB were carried out in India and the fabrication of other components and assembly were done at CERN.

\subsection{GEM foils}
\begin{figure}[h]
	\centering
 \includegraphics[height=0.25\textwidth,width=5cm,angle=0]{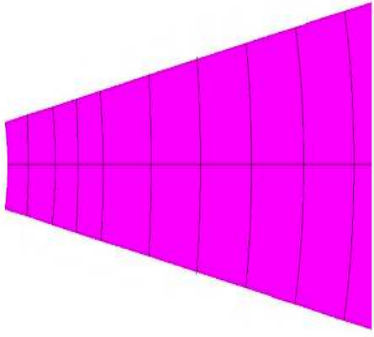}
  \caption{Layout of the HV segmentation on the GEM foil, only a part of the foil with only 20 segments have been shown.}
\label{large-gem-foil-segmentation}
\end{figure}

This prototype triple GEM chamber is made of 3 standard single mask GEM foils. The drift gap, transfer gap and the induction gap of 
the chamber are $3~mm, 1~mm,1.5~mm$ respectively. The GEM foils for the prototype chamber have been fabricated at the CERN. The 
GEM foils have the provision of stretching by 
NS-2 (no stretch, no spacer) technique\cite{NS2}. The layout of the high voltage segmentation on the foil is shown in Fig.~\ref{large-gem-foil-segmentation}. The segmentation 
is made based on the occupancy of the chambers in Au$+$Au collisions at SIS-300 energy. Therefore, it is expected that the chamber will 
be able to handle particle rate at lower energy collisions available at SIS100 quite comfortably. 
Each GEM foil has been segmented into 24 sections on its upper surface. The innermost four sections were of $25~ cm^2$, rest are $100 ~cm^2$ area. 
Each of 24 sections was connected via a surface-mounted $1~M\Omega$ protection resistance. Four zones each having 6 segments were 
connected to independent power supplies using four resistive chains. For the $100~ cm^2$ area with a $1~M\Omega$ protective resistance, 
calculation shows a voltage drop of $0.4~V$ due to a pulse current for a particle rate of $10~MHz$. This drop in voltage does 
not change gain significantly.

     \begin{figure}
      \begin{center}
 	\includegraphics[width=50mm, height = 85 mm]{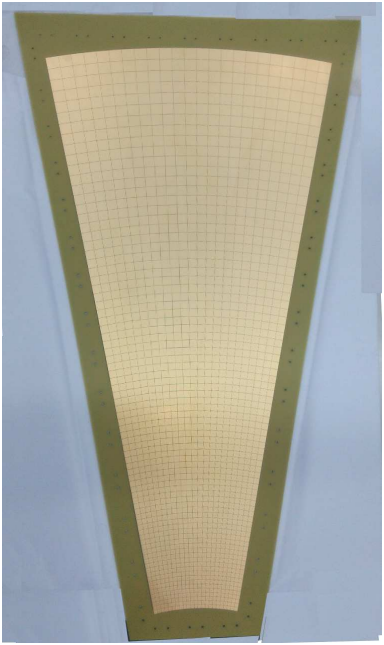}
      \end{center}
	\caption{Sector shaped readout PCB.}
	\label{fig:chamber}
  \end{figure}

\subsubsection{Drift Plane} 
The drift plane is a $3~mm$ thick plane with copper clad on single side, fabricated at the CERN as per design from VECC. 
The drift printed circuit board (PCB) was extended laterally by $5~mm$ in order to accommodate the HV lines for powering the segments. 14 holes each of $2~mm$ diameter are 
made at appropriate positions to allow X-ray to pass through during testing. The holes are covered with mylar foils to make the chamber gas-tight.
% We have shown the drift plane the HV divider connections in Fig.~\ref{large-chamber-hv-divider}.

\subsection{Readout Plane}
The anode readout PCB is an eight layered, $2.7~mm$ thick PCB designed at VECC and fabricated in Bangalore, India.
 The PCB has $1^\circ$ progressive size readout pads as shown in Fig.~\ref{fig:chamber}. The angle has been arrived at after simulations as mentioned in \cite{sim}. 
 Each pad is a trapezium whose larger angles are $90.5^\circ$. This PCB has 23 pads in an annular ring and 79 pads 
 in the radial direction. The dimension of the readout pads are from $3.9~mm \times 3.9~mm$ to maximum $16.6~mm \times 16.6~mm$. This way multi-hit 
 probability on a pad is reduced. The coarse granularity of readout pads in the outermost region also drastically reduce the cost of the readout electronics. 
 In total there are 1817 pads in the anode PCB.

\subsubsection{Assembly of the chamber}

\begin{figure}
\centering
 \includegraphics[width=70mm, height = 80 mm]{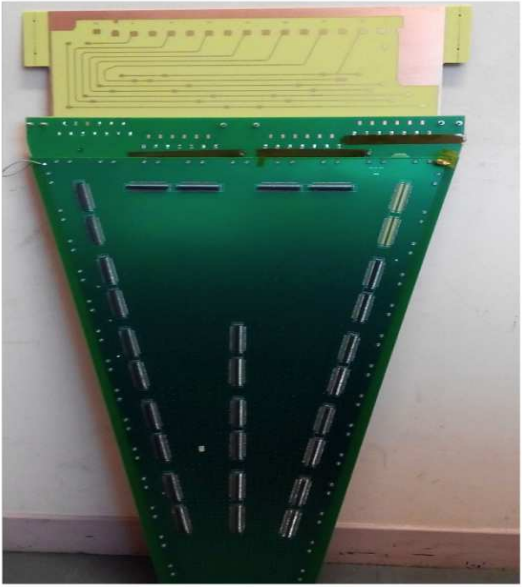}
 \caption{An assembled sector-shaped real-size chamber for the first station of the CBM muon chamber.}
 \label{fig:ready_chamber}
\end{figure}

Some of the challenges that require special care for building the real size chamber are : (a) building of a large size chamber PCB (b) fabrication of a large size 
GEM foil (c) stretching of the large size foils and (d) proper layout of the tracks to accommodate the variation of occupancy. 
The job has been performed keeping in close contact with the RD51 team at CERN. 
\begin{figure}[h]
	\centering
 	\includegraphics[width=0.45\textwidth,height=10cm]{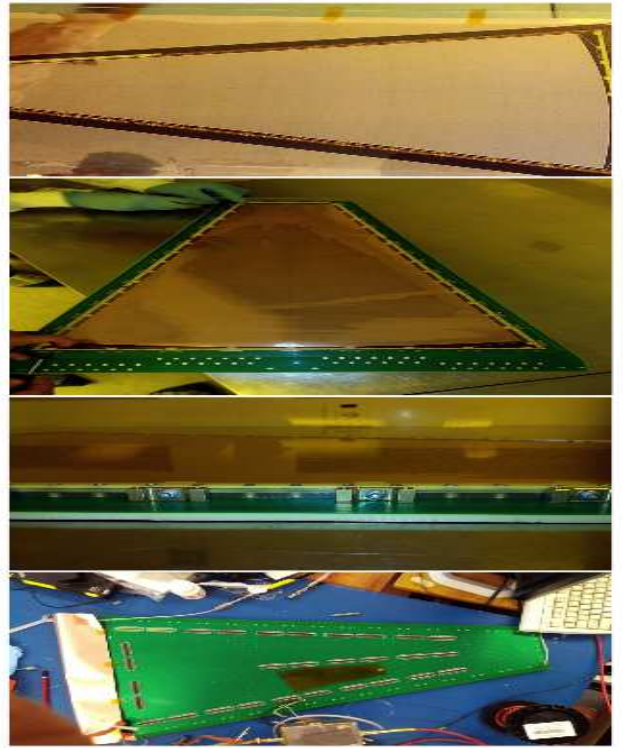}
  \caption{GEM foil stretching using NS-2 technique, (top-left) special screws connected to spacers on the edge of the chamber, (top-right) A chamber ready for NS-2 stretching, (bottom-left) a GEM foil stretched and assembled, (bottom-right) view of several layers.}
\label{ns2-combo}
\end{figure}
We have used the NS-2 technique developed at CERN which has the advantages that (i) foils can be easily replaced 
(ii) no permanent gluing or thermal stretching is done and the foil could be reused, if required. The assembly of the chamber using NS-2 technique is shown 
in Fig.~\ref{ns2-combo}. First, brass pieces are fixed at the boundaries of the drift plane at a regular interval (Fig.~\ref{ns2-combo}.(top)). They act as support pillars with holes 
at prescribed intervals against which the foils will be stretched. A $1~mm$ G10 spacer frame is placed between two foils to provide $1~mm$ transfer gap. 
Thin metallic pins of appropriate size are soldered on the drift plane and passing through the spacers make contact with the respective GEM foils. 
Next the readout plane is placed keeping $1.5~mm$ induction gap. The entire chamber is sealed by the 
anode readout plane via an O ring. The screw pins are tighten to stretch the foils. Now the chamber is cleaned in an ultrasonic bath for some minutes. 
Finally, the chamber is made dry using an oven and put under $Ar:CO_{2}$ gas. The Fig.~\ref{fig:ready_chamber} shows an assembled sector-shaped chamber showing the 
connecting points for the HV at the top attachment. All the design works were performed at VECC-Kolkata and the assembly of the large GEM chamber was 
performed at CERN.

 \section{Test Beam Setup}
\begin{figure}[h]
    \begin{center}
	\includegraphics[width=85mm, height = 50 mm]{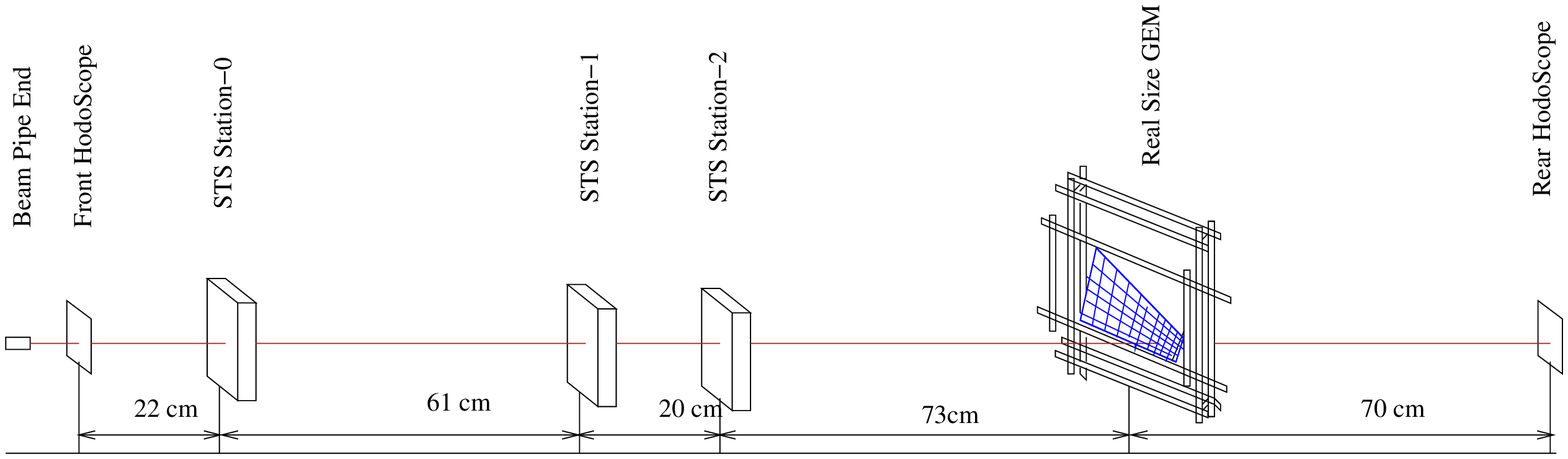}
    \end{center}
	\caption{Experimental Setup at COSY, J\"{u}elich, Germany.}
	\label{fig:setup}
  \end{figure}

 A schematic layout of the experimental setup for testing the chamber is shown in Fig.~\ref{fig:setup}. The GEM chamber was tested along with three 
 Silicon Tracking Stations (STS) at the Jessica cave of COSY, J\"{u}elich, Germany using proton beam of momentum $2.36~GeV/c$. A pair of crossed optical 
 fiber scintillators hodoscopes with the overlapping area of  $2~cm \times 2~cm$ placed at two ends of the setup has been used to form the beam trigger. 
 The coincidence signals from the front and rear hodoscope scintillators were connected to the ReadOut Controller (ROC) for 
 recording the timestamps of the beam particles. 
 The detectors in this setup (STS, GEM, hodoscopes) were read out by using the nXYTER front end boards (FEB) directly connected to the chamber 
 followed by the Readout controller (ROC). One ROC can handle two FEE boards. The readout PCB was divided into 15 regions, each one is read by one 
 nXYTER that has 128 channels connected to 128 readout pads. nXYTER has one fast channel and another slow channel of peaking times
 $30~nanosecond$ and $140~ nanosecond$ respectively and has a 12-bit ADC of $25fC$ dynamic range. Data were collected in trigger-less condition or in self-triggering mode. The nXYTER ASIC 
 records the time-stamps of each hit on the detector above a predefined threshold. The time-stamps of all the hits above threshold are digitized 
 and stored. Data were recorded for different beam intensities by adjusting the collimator windows. Data at different voltage-settings across the GEM foils
 were taken for different regions of the detector where readout pads of different sizes were exposed to the beam. All GEM 
 foils in the chamber were kept at same voltage setting. A premixed gas mixture of argon (Ar) and carbon dioxide (CO$_{2}$), mixed in 70:30 ratio by mass, was used. 

 \section{Results}
 \begin{figure}[!h]% time correlation
      \begin{center}
		\includegraphics[width=80mm, height = 50 mm]{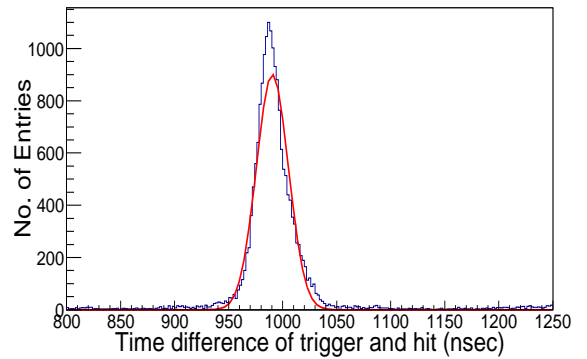}
      \end{center}
      \caption{Distribution of the time difference between time-stamps of the trigger and GEM signals.}
% 	\caption{Time correlation spectra of aux signals and hits on GEM detector}
	\label{fig:timecorrelation}
  \end{figure}
\begin{figure}[h]
      \begin{center}
	\includegraphics[width=80mm, height = 50 mm]{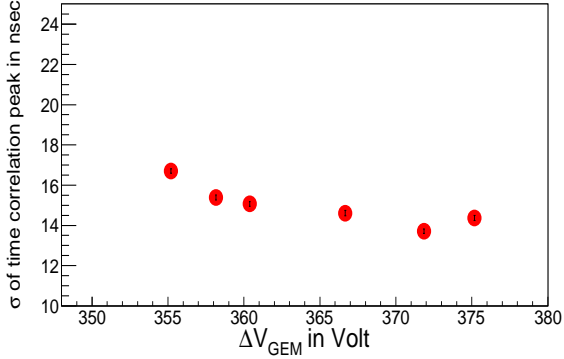}
      \end{center}
	\caption{Variation of sigma of time correlation with $\Delta V_{GEM}$.}
	\label{fig:sigma_volt}
  \end{figure}
\begin{figure}[h]
      \begin{center}
	\includegraphics[width=80mm, height = 50 mm]{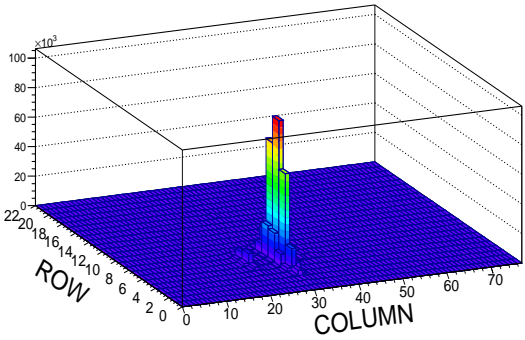}
      \end{center}
	\caption{Proton beam spot on the GEM chamber in the region of pad size $7.39~mm\times7.39~mm$.}
	\label{fig:beamspot1}
  \end{figure}

  In a self-triggered readout system where all hits are stored along with their timestamps, first step of data analysis would be to find hits that 
  are correlated in time with the trigger time stamps. A distribution of the difference in timestamps between coincidence trigger signal 
  from the hodoscopes and those of the hits are shown in Fig.~\ref{fig:timecorrelation}. The time correlation 
 distribution is fitted with a Gaussian with mean = $990.2~nanosecond$ and $\sigma$ = $13.71~nanosecond$ at a voltage across the GEM foil $(\Delta V_{GEM})$of$~=~371.8~V$. 
 The position of the peak depends on the cable delay in addition to the delay introduced by electronics.
 Almost no entries ($\sim~3-4\%$) outside the peak region suggest that most of the hits 
 are correlated with the trigger. The $\sigma$ of the peak is a measure of the time resolution of the detector. The variation of $\sigma$ with  
 $\Delta V_{GEM}$ at a fixed location is shown in Fig.~\ref{fig:sigma_volt}. The variation has a minimum at $13.71~nanosecond$ at $\Delta V_{GEM} = ~371.8~V$.
 Hits lying inside the time correlation peak are related to beam particles and hence considered for further analysis. The positions 
 of such hits on the GEM is shown in Fig.~\ref{fig:beamspot1} for the region where the pad size was $7.39~mm\times7.39~mm$. We 
 get a narrow beam-spot for proton beam. The beam is mainly confined within a few pads.

 In the next subsection, we will discuss different results on the chamber properties like ADC distribution, 
 gain, efficiency among others in detail.

 \subsection{ADC distribution and chamber gain } 
\begin{figure}[!h]
      \begin{center}
\includegraphics[width=80 mm, height = 50 mm]{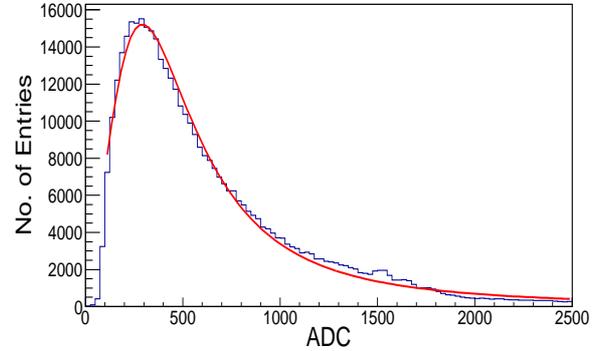}% new 2

      \end{center}
	\caption{Pedestal subtracted cluster ADC distribution of the GEM chamber.}
	\label{fig:adc}
  \end{figure}

\begin{figure}[!h] % gain vs voltage
      \begin{center}

		\includegraphics[width=80mm, height = 50 mm]{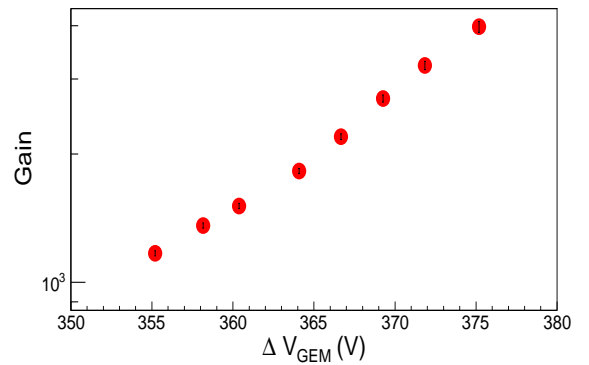} %new

      \end{center}
      	\caption{Variation of chamber gain with $\Delta V_{GEM}$.}
%	\caption{Variation of MPV of the adc spectra taking cluster of $3\times3$ cells around the beam spot}
	\label{fig:cluster}
  \end{figure}

\begin{figure}[!h] % gain uniformity channel
      \begin{center}
		\includegraphics[width=80mm, height = 50 mm]{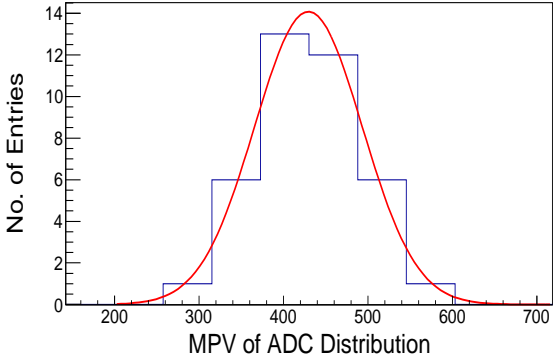}
		
      \end{center}
      \vspace{2mm}
      	\caption{Uniformity of MPV of adc distribution for different channels with mean = 429.6 adc and sigma = 63.81 at  $\Delta V_{GEM}~=~369.3~V$.}
	\label{fig:Guni}
  \end{figure}

 ADC values of the hits within the time correlation window in an event are summed up to obtain the total ADC of a cluster in an event. 
 The event-by-event pedestal subtracted ADC distribution is shown in 
 Fig.~\ref{fig:adc} for $\Delta V_{GEM} = ~366.7~V$. The distribution is fitted with a Landau distribution with MPV $=~ 386.6$.
% The event by event pedestal subtracted ADC spectra are plotted for different $\Delta V_{GEM}$ for the events with single hit on GEM. One of the ADC distribution
 %at $\Delta V_{GEM} = ~366.67~V$ fitted with Landau distribution is shown in Fig.~\ref{fig:adc} with MPV = 383.7.
 The MPV values of ADC distributions are calculated for different $\Delta V_{GEM}$ on the GEM foil. The ADC distribution saturates at higher $\Delta V_{GEM}$ due to 
 limited dynamic range of the nXYTER. The MPV values are used to calculate the total charge collected by the chamber. The input charge is the 
 charge of electrons created by primary ionization in a $3~mm$ drift gap. The number of primary electrons is taken to be 30. 
 The gain of the triple GEM prototype rises linearly with voltage across each GEM foil as seen in Fig.~\ref{fig:cluster}. Pad by pad variation 
 of gain is shown in Fig.~\ref{fig:Guni} for 39 pads of varying dimensions from $4.44~mm \times 4.44~mm$ to $5.61~mm \times 5.61~mm$. 
 It is observed that the gain is reasonably uniform over 
 the entire detector plane with a standard deviation of $12\%$.
 %This ensures that total charge to be collected by a pad and the effect of varying pad size can be studied. 

\begin{figure}[!h] % gain vs rate
      \begin{center}
		\includegraphics[width=80mm, height = 50 mm]{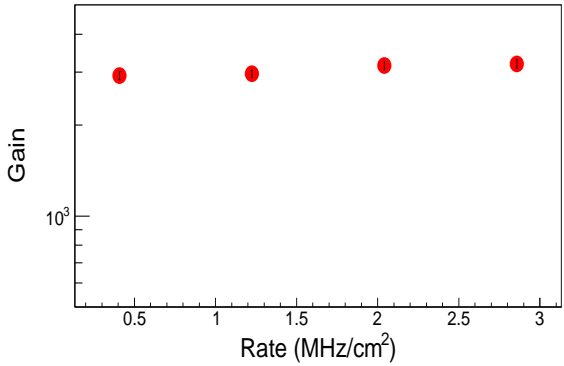} %new

      \end{center}
      	\caption{Stability of gain with the rate of beam particles.}
	\label{fig:gain_rate}
  \end{figure}

 We also estimate the stability of the gain because of increase of the rate of the incoming particles.  The rate has been obtained by the difference between the average 
 time intervals between two consecutive sets of 100 trigger signals. The gain is almost stable with a variation $\sim~9\%$  at the highest rate of $2.8~MHz/cm^{2}$
 as shown in Fig.~\ref{fig:gain_rate}.
 
     \begin{figure} % cell mult uniformity
      \begin{center}
		\includegraphics[width=80mm, height = 50 mm]{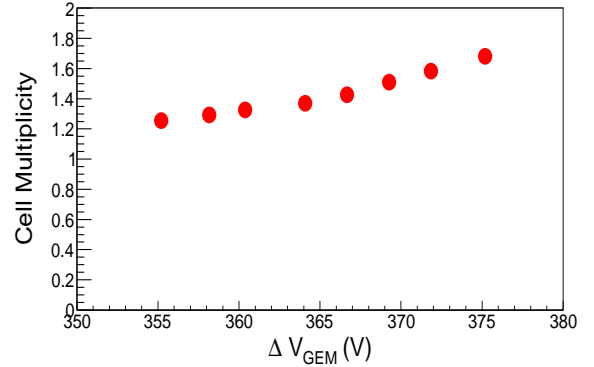}

      \end{center}
	\caption{Variation of cell multiplicity with $\Delta V_{GEM}$, the size of the error bars are smaller than symbol size.}
	\label{fig:cellmult_vol}
  \end{figure}
  
 \subsection{Cell Multiplicity}
 Cell multiplicity is calculated using the hits within the selected time correlation window. 
 The average cell multiplicity is slowly increasing with $\Delta V_{GEM}$
 from 1.2 at $\Delta V_{GEM}~=~355.2~V$ to 1.6 at $\Delta V_{GEM}~=~375.2~V$ as can be seen from Fig.~\ref{fig:cellmult_vol}. 
 Due to the increase in voltage across GEM, the gain increases resulting in an increase of transverse size of the cluster profile.
 
   \begin{figure} % eff with vol
      \begin{center}
		\includegraphics[width=80mm, height = 50 mm]{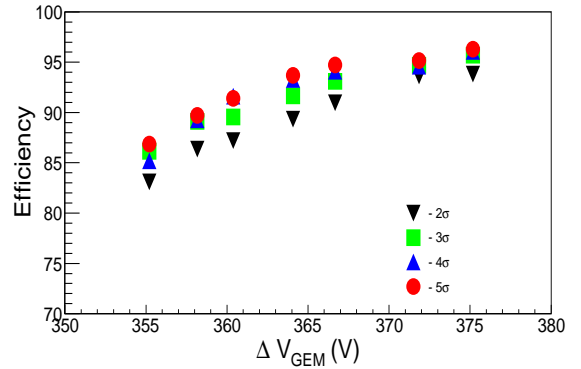}
      \end{center}
	\caption{Efficiency with $\Delta V_{GEM}$ for different time correlation distribution window, the size of the error bars are smaller than the symbol size.}
	\label{fig:eff_volt}
  \end{figure}

 \subsection{Efficiency}
\begin{figure} % eff uniformity
      \begin{center}

 		\includegraphics[width=80mm, height = 50 mm]{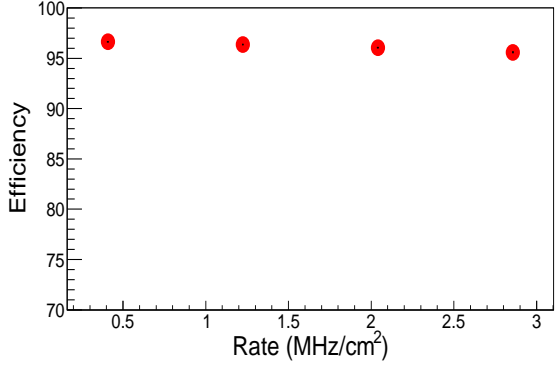} %new

      \end{center}
	\caption{Variation of efficiency with rate of incoming particle.}
	\label{fig:eff_rate}
  \end{figure}

 The GEM detector will be used for tracking muons in CBM, so to operate the GEM detector in CBM-MUCH at high interaction rate 
 the efficiency of the detector should be $>95\%$.
 As the GEM detector is aligned with the hodoscopes, the particles that have correlated hits on both the front and rear hodoscopes are taken as the 
 input particle on to the GEM. Particles are said to be detected if it has at least one hit on the GEM chamber within the time correlation window. 
 The ratio of the number of detected particles as defined above and the number of triggers in a given time interval gives the efficiency 
of the detector. The variation of efficiency with $\Delta V_{GEM}$ is shown for time windows of $~2~\sigma,~3~\sigma,~4~\sigma,~5~\sigma$ of the time correlation spectra.
 The study of the window size shows that a size of $3\sigma$ seems optimum. The efficiency increases with $\Delta V_{GEM}$ of the detector 
 and reaches $96\%$ at $\Delta V_{GEM}= 375.2~V$ for time windows equal to $3\sigma $ as shown in Fig.~\ref{fig:eff_volt}. MUCH will be used in 
 high interaction rate so efficiency should be stable for high rate of beam particles.  
The variation of efficiency with the average rate of the incoming particles 
 is shown in Fig.~\ref{fig:eff_rate}. The efficiency at a fixed $\Delta V_{GEM} = 375.2~V$ is shown upto a beam rate of $2.8~MHz/cm^{2}$. The size of the beam 
 is determined from its spot size on the hodoscopes. This rate is higher than the maximum particle-rate that the first MUCH detector has to 
 face. As the particle rate increases efficiency decreases slightly ($\sim~2\%$).
%   The variation of efficiency with particle-rate is shown in Fig.~\ref{fig:eff_rate}.

 \subsection{Summary and discussions}
 The main challenge of MUCH detector is to handle high rate of incident particles.
This is the first report on the performance of a real-size large GEM chamber suitable for the first MUCH chamber for its rate capability.
A premixed $Ar:CO_{2}$ gas mixture in $70:30$ ratio by mass is used. The detector was readout in self-triggered mode using the nXYTER. The hits on GEM detector are 
correlated in time to the signals produced by a pair of crossed scintillators hodoscopes one at the front and another at the rear 
position. The time correlation 
distribution is fitted by a Gaussian distribution with $\sigma =~13.71~nanosecond$ at $\Delta V_{GEM}~=~371.9~V$. $\sigma$ is related to the time resolution of the detector. The efficiency 
of the detector reaches $96\%$ at $\Delta V_{GEM}~=~375.2~V$. Cell multiplicity at this voltage for proton beam is $1.6$. Cell multiplicity increases with 
$\Delta V_{GEM}$ because the transverse size of the beam increases. The efficiency of the detector slightly decreases due to increase of the rate of the particle but the 
change is $\sim 2\%$ ( $96.0\%$ at $0.4~MHz/cm^{2}$ to $94.8\%$ at $2.8~MHz/cm^{2}$ average particle-rate for $\Delta V_{GEM}~=~375.2~V$).
%The  of the detector does not change too much 
%due to high rate( $96\%$ at $75~kHz$ to $93.4\%$ at $800~kHz$ average particle rate for $\Delta V_{GEM}~=~371.9~V$). 
 \section{Acknowledgement}
 We thank S. K. Ghosh of Bose Institute, Kolkata, Walter Mueller, P. Senger of GSI-Darmstadt, L. Ropelski and E. Oliveri of RD$51$ for all help. We would 
 also like to thank the crew of COSY accelerator at Juelich, Germany. This work is supported by the DAE-SRC award under the 
 scheme no. $2008/21/07-BRNS/2738$. The work has been funded by Department of Atomic Energy, Government of India and the Department of Science 
 and Technology, Government of India. RP and SS thank two Indian funding agencies, University Grants Commission and Council of Scientific and Industrial
Research for their grants with reference numbers F. No. $~2-8/2002~(SA-I)$ and $09/015(0397)/2010-EMR-I$ respectively.

 % reference bibliographic

\end{document}